\documentclass[useAMS,usenatbib]{mn2e}
\usepackage{graphicx}

\newcommand{\apg}{\ga}
\newcommand{\apll}{\la}

\newcommand{\hmsol}{\mbox{$h^{-1}\,{\rm M}_\odot$}}
\providecommand{\kms}{\,\ensuremath{\rm{km\,s}^{-1}}}
\newcommand{\hkpc}{\mbox{$h^{-1}~{\rm kpc}~$}}

\title[The Star Formation History of Luminous Red Galaxies Hosting Mg\,II Absorbers]{The Star Formation History of Luminous Red Galaxies Hosting Mg\,II Absorbers}

\author[Jean-Ren\'e Gauthier and Hsiao-Wen Chen]{Jean-Ren\'e Gauthier\thanks{E-mail:
gauthier@oddjob.uchicago.edu (JRG); hchen@oddjob.uchicago.edu (HWC)} and Hsiao-Wen Chen\footnotemark[1]\thanks{This paper includes data gathered with the 2.5 meter du Pont telescope located at Las Campanas 
Observatory, Chile and with the Apache Point Observatory 3.5-meter telescope, which is owned and operated 
by the Astrophysical Research Consortium. }\\
Department of Astronomy \& Astrophysics and Kavli Institute for Cosmological Physics, University of Chicago, IL, 60637 USA}
\begin{document}

\date{Accepted 2011 August 18. Received 2011 August 17; in original form 2011 July 21}

\pagerange{\pageref{firstpage}--\pageref{lastpage}} \pubyear{2011}

\maketitle

\label{firstpage}

\begin{abstract}
  We present a spectroscopic sample of $z \approx 0.5$ luminous red
  galaxies (LRGs) that are located within physical projected distances
  $\rho \la 350$ \hkpc of a QSO sightline. Of the 37 LRGs in our
  sample, eight have associated Mg\,II absorbers with rest-frame
  equivalent width $W_r(2796)>0.3\,$\AA\ and velocity separation
  $|\Delta v| \la 350$ \kms\ and 29 do not have associated Mg\,II
  absorbers to a 2-$\sigma$ limit of $W_r(2796)=0.3\,$\AA.  We perform
  a stellar population synthesis analysis using stacked spectra of the
  Mg\,II absorbing and non-absorbing LRG subsamples.  We find that
  LRGs with or without associated Mg\,II absorbers share similar star
  formation histories and are best described by old stellar population
  models ($\ga$ 1 Gyr).  Younger stellar populations ($\la$ 1
  Gyr) fail to reproduce their spectra. These findings are consistent
  with the lack of [O\,II] emission features in the LRG spectra.  The
  primarily old stellar populations in the LRGs indicate that
  starburst driven outflows are unlikely to explain the observed
  Mg\,II absorbers at large distances from the LRGs.  In addition, the
  spectroscopic LRG sample allows us to derive a sensitive constraint
  for the cool gas covering fraction of $\langle\kappa\rangle=14\pm
  6$\% in the LRG halos for absorbers of $W_r(2796)>0.3\,$\AA.
  Finally, we speculate on the origin of the observed Mg\,II absorbers
  around the LRGs.
\end{abstract}


\section{Introduction}

Tracing the origin of the baryonic content of galaxies is a key
ingredient to a comprehensive theory of galaxy formation. In Gauthier
et al. (2009,2010; hereafter G09,G10), we studied the cool gas content
of luminous red galaxies (LRGs) using a sample of close QSO--LRG
pairs.  Cool gas is revealed by the presence of Mg\,II
$\lambda\lambda$ 2796,2803 absorption systems in the spectra of
background QSOs at the redshifts of the LRGs (see also
\citealt{bowen2011a}). Our studies revealed a non-negligible amount of
cool gas with covering fraction $\kappa \approx 20\%$ in LRG
halos. The association between LRGs and Mg\,II absorbers is
unexpected, because LRGs inhabit dark matter halos of mass $\sim
10^{13}$ \hmsol (e.g.,\ \citealt{blake2008a}) for which theoretical
models and hydrodynamical simulations predict that little cool gas can
survive (e.g.,\
\citealt{birnboim2003a,keres2009a,stewart2010a,faucher-giguere2011a,mo1996a,maller2004a}).
In addition, LRGs constitute a homogeneous sample of passive galaxies
characterized by old stellar populations giving rise to strong 4000-\AA\ breaks. 
They are characterized by luminosities of $\approx 5$ $L_*$ and stellar masses of 
$\approx 3 \times 10^{11}~{\rm M}_\odot$ at $z\approx0.5$ (e.g.,\ \citealt{tojeiro2011a}). Given 
the relatively old stellar populations of these galaxies, starburst-driven
outflows are unlikely to produce the observed cool gas in LRG halos.

But while most LRGs are quiescent and show little star formation
activity, \citet{roseboom2006a} showed that $\approx$ 10\% of the
galaxies display [O\,II] $\lambda\lambda$ 3727,3729.  It is possible
that the LRGs hosting Mg\,II absorbers are a biased sub-population
that contain on-going star formation.
In this \emph{letter}, we present a stellar population synthesis
analysis of LRGs spectroscopically identified at physical projected
separations $ \leq 350$ \hkpc from a QSO sightline.  Comparing the
stellar populations and star formation histories of Mg\,II absorbing
and non-absorbing LRGs allows us to examine whether or not Mg\,II
absorbers occurs preferentially in the star-forming subsample.  In
addition, the spectroscopic LRG sample enables us to constrain the
incidence of Mg\,II absorbers in LRG halos based on inspections of the
QSO spectra.

\begin{centering}
\begin{figure}
 \vspace{0.5pt}  \centerline{\hbox{ \hspace{0.0in}
	\includegraphics[angle=0,scale=0.30]{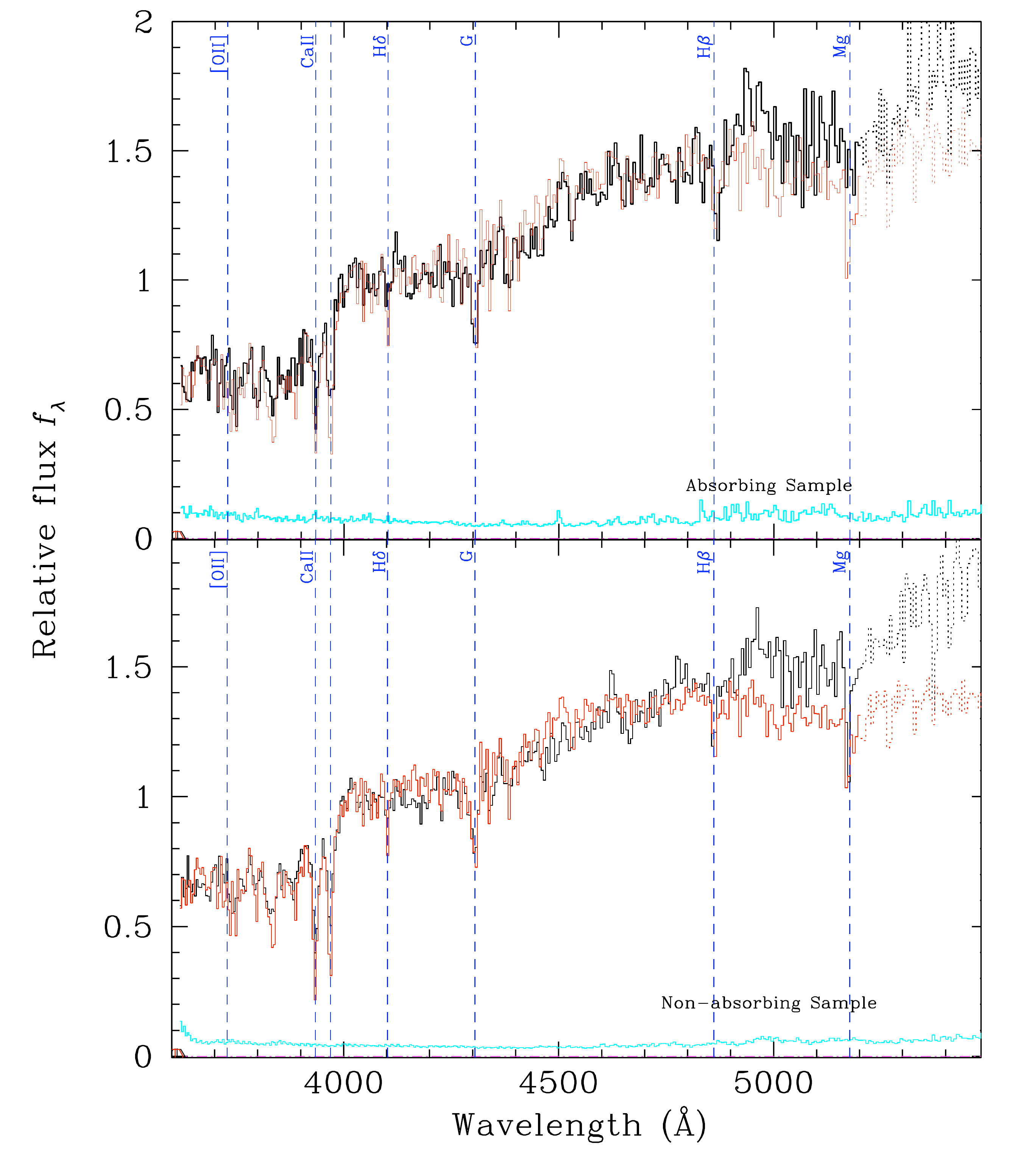}
      }
   }
\caption{Comparisons of the observed LRG spectra and best-fit models from 
a stellar population synthesis analysis. The top panel shows the stacked spectrum 
of eight LRGs with associated Mg\,II absorbers at projected distances $<$ 350 \hkpc\, along 
with the best-fit synthetic model displayed in red. The data points and associated model 
predictions shown in dotted lines have been excluded from the stellar population synthesis analysis. 
The bottom panel shows the stacked 
spectrum of 29 LRGs without associated Mg\,II absorbers down to a 2-$\sigma$ limit 
of $W_r(2796)=0.3$\AA\ , in comparison to the best-fit model displayed in red. 
Spectral features such as Ca\,II H\&K, G-band, and Mg\,I are prominent in both 
stacked spectra, while [O\,II] emission features are not seen. The 1-$\sigma$ error spectrum 
of each stack is shown at the bottom of each panel. }
\label{best}
\end{figure} 
\end{centering}

\section{Observations and Data analysis}

We conducted long-slit spectroscopy of 37 LRGs at projected
separations $\rho \leq 350$ \hkpc from background QSO sightlines. The
LRGs were photometrically identified in SDSS DR4
\citep{collister2007a,blake2007a} and the QSOs were selected from the
SDSS DR5 QSO spectroscopic catalog \citep{schneider2007a}. The maximum
separation corresponds approximately to the virial radii of the LRGs
(see G09), allowing us to probe gravitationally bound gas.  

The LRGs were separated into two subsamples based on the prior knowledge 
of the presence of Mg\,II absorbers in the QSO spectra. In Sample A, photometrically-selected 
LRGs were found in the vicinity of \emph{known} Mg\,II absorbers. In Sample B, the LRG--QSO pairs were 
composed of random LRGs near QSO sightlines for which we had no prior knowledge of the presence or 
absence of Mg\,II absorbers. Sample B was used to derive constraints on the 
covering fraction of cool gas. Of the 37 LRGs presented in this paper, 28 belong to 
Sample B and nine to Sample A. More details about the samples can be found in G10. 

We used
the Double Imaging Spectrograph (DIS) on the 3.5-m telescope at the
Apache Point Observatory and the Boller \& Chivens spectrograph (B\&C)
on the du Pont telescope at the Las Campanas Observatory in
Chile. Details about the spectroscopic observations and data reduction
are presented in G10.  In summary, we acquired long-slit galaxy
spectra at intermediate resolution ($R \approx 1000$) in the
wavelength range 5000-8500 \AA.  The raw spectra were flux-calibrated
using a spectrophotometric standard.  To examine the accuracy of the
flux calibration, we compared the spectral energy distribution of our
flux calibrated LRG spectra with available $g'r'i'z'$ broad-band
photometry from the SDSS data archive.  We confirmed that our flux
calibration yielded consistent continuum slopes for the LRGs at
wavelengths $\lambda \apll 8000$ \AA.

We determined the redshifts of the LRGs using a cross-correlation analysis
with known SDSS galaxy templates. Typical redshift uncertainties were
$|\Delta v| \approx 70$ \kms at $z \approx 0.5$. All galaxy spectra
were corrected for atmospheric and Galactic extinction.  A journal of
the spectroscopic observations are presented in columns (1)-(6) of
Table 1.

At the spectroscopic redshift of the LRGs, we inspected the QSO
spectra and searched for Mg\,II absorbers within $|\Delta v| = 350$
\kms\ .  This velocity interval corresponds roughly to the virial
velocity of the LRGs. When a Mg\,II absorber is found in the QSO
spectrum, we measured the redshift and strength of the absorber using
a Gaussian profile analysis. When no absorber was visually detected,
we classify the LRG as a non-absorbing galaxy. Column (7) of Table 1
displays the Mg\,II absorption strength.  In cases where no Mg\,II
features are detected, we present a 2-$\sigma$ upper limit of the
absorber strength. Two galaxies (SDSSJ114445.36,SDSSJ220703.36) in the
absorbing sample have velocity separations of 380 and 360 \kms, which
are larger than the considered limit $|\Delta v| = 350$ \kms\ . Given
that the redshift uncertainty of the galaxies is $|\Delta v| \approx
70 \kms$, we cannot rule out the possibility that the absorbers are
associated with the LRGs. We therefore included them in the absorbing
sample.  We tested whether assigning these galaxies to the absorbing
sample altered the results of the population synthesis analysis or not
and we found no changes.  In summary, the spectroscopic observations
of the LRGs yielded eight physical LRG-Mg\,II pairs and 29 LRGs
without associated Mg\,II absorbers to a sensitive upper limit.  Note
that we retrieved the spectrum of one LRG (SDSSJ125300.00) from the
SDSS archive and included it in our sample. Projected separations and
velocity separation of the LRG--QSO pairs are presented in Columns
(8)-(9) along with the LRG sample (A or B) in column (10) of Table 1.

Individual spectra of the LRGs in our sample have typical $S/N \approx
4$.  To obtain robust results from the population synthesis analysis,
we formed a stacked spectrum of each of the LRG subsamples. To
generate a stacked spectrum, we first masked out strong sky emission
lines and absorption bands.  We then shifted the observed spectrum to
the rest-frame of the galaxy and adopted a common pixel resolution of
$\Delta v$=350 \kms. We employed a weighting scheme in which each
galaxy spectrum is weighted by their mean $(S/N)^2$ to maximize the
final $S/N$.  The stacked spectrum of each sample covers the
rest-frame wavelength range 3600--5500\AA\ and is shown in Figure 1.

\section{Stellar Population Synthesis Analysis}

We performed a stellar population synthesis analysis to constrain the
ages of the stellar populations of the LRGs.  To accomplish this task,
we carried out a likelihood analysis that compares the stacked
spectrum with model expectations for different stellar age,
metallicity and star formation history. The likelihood function is
defined as
\begin{equation}
\mathcal{L}(t) = \prod_{i=1}^N \exp \Bigg \{ -\frac{1}{2} \Bigg[ \frac{f_i - \bar{f}_i(t)}{\sigma_i}   \Bigg]^2  \Bigg \}
\end{equation}
where $t$ is the age of the stellar populations, $N$ is the number of
spectral bins ($N=311$), $f_i$ is the observed flux in the $i$th bin,
$\bar{f}_i$ is the model prediction, and $\sigma_i$ is the
corresponding error of the $i$th element.  As described in \S\ 2, the
flux calibration became uncertain at $\lambda\apg 8000$ \AA,
corresponding to rest-frame $\lambda\apg 5200$ \AA\ for these LRGs.  We
therefore limited our analysis to the spectral range 3600--5200 \AA\
in the rest-frame of the LRGs (Figure 2).


The stellar population models were based on those described in
\citet{bruzual2003a} revised to include a prescription of the TP-AGB
evolution of low and intermediate mass stars \citep{marigo2007a}.  We
employed a Chabrier initial mass function for all models
\citep{chabrier2003a}.  The star formation history (SFH) of the model
galaxies was parametrized by either a single burst or by an
exponentially declining model with an $e$-folding timescale $\tau$.
The ages $t$ were equally separated in logarithmic space between
$10^5$ yr and 8.4 Gyr, where the upper-limit corresponds to the age of
the Universe at $z \approx 0.5$.  We also adopted an equal spacing of
50 Myr for $\tau$ from 0.1 to 0.5 Gyr and we adopted metallicities of
0.02, 0.2, 1, and 2.5 solar.  To directly compare between data and
models, we convolved the model spectra with a top-hat function of
width 350 \kms\ to match the resolution of the data.

Extinction by the host LRG was not included in the final model templates.  
However, we did generate a library of model spectra with host extinction 
following the Charlot \& Fall (2000) prescription. We found that the LRG 
spectra systematically select the models with the least amount of extinction.
For this reason, we did not include extinction by the host galaxy 
in the final library of synthetic spectra.

\begin{figure}
 \vspace{0.5pt}  \centerline{\hbox{ \hspace{0.0in}
	\includegraphics[angle=0,scale=0.35]{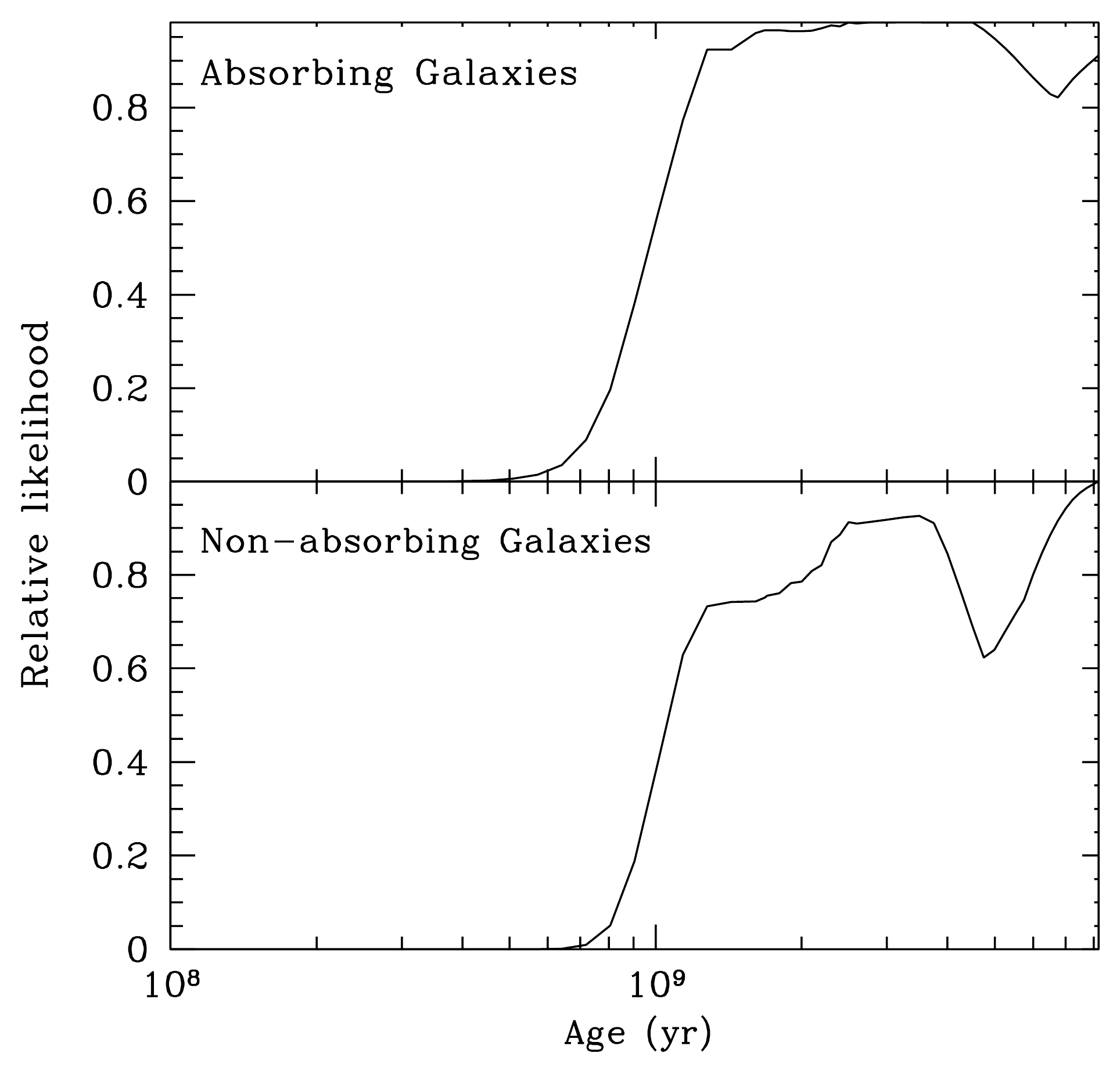}
      }
   }
\caption{Relative likelihood functions of the stellar age of LRGs with and without associated Mg\,II absorbers. 
The stellar age distribution shows that both galaxy samples are characterized by an evolved stellar population 
of age $\ga$ 1 Gyr. The model spectra are shown in Figure 1. 
The Mg\,II absorbing LRG sample is best characterized by a
$\tau$ model of $\tau=0.15$ Gyr, age $t=3.25$ Gyr, and solar
metallicity, while the non-absorbing LRGs are best-characterized by a
single burst of age $t=8.25$ Gyr and metallicity $0.2$ solar.
At $\la 1.6$ Gyr, the LRGs can be characterized by a single burst of metallicity 2.5 solar. 
At $1.6 \la t \la 5$ Gyr, the LRGs can be characterized by an
exponentially declining SFR model of $\tau=0.2$ to 0.5 Gyr and metallicity of solar or 2.5 solar.
At older ages ($\ga 5$ Gyr), the best-fit models are either single burst of 
subsolar metallicity (0.2 solar) or exponentially declining models of $\tau=0.5$ Gyr and solar metallicity.  
}
\label{likelihood}
\end{figure}

\section{Mean Stellar Population of the LRGs}

Figure \ref{best} shows the stacked spectra of both samples of LRGs
along with the best-fit stellar population model in red.  The LRGs in
both samples exhibit spectral features dominated by absorption
transitions, suggesting an old underlying stellar population and
little star formation in the recent past.  
Visual inspections of individual LRG spectra show that three LRGs,
(SDSSJ142312.00, SDSSJ220703.36, SDSSJ232924.13) exhibit traces of
[O\,II] emission, one of which belongs to the absorbing sample
(SDSSJ220703.36). These three LRGs represent 8\% of the absorbing and
non-absorbing LRGs combined. The fraction of [O\,II] emitting LRGs is
consistent with the finding of Roseboom et al. (2006), who found that
$\sim$10\% of LRGs show [O\,II] emission.  

The spectra of the LRGs support the previous understanding that these
galaxies are primarily quiescent, which is further supported by the
results of the population synthesis analysis presented in Figure 2.
The relative likelihood functions of the stellar age of these LRGs
show that the Mg\,II absorbing LRG sample is best characterized by a
$\tau$ model of $\tau=0.15$ Gyr, age $t=3.25$ Gyr, and solar
metallicity, while the non-absorbing LRGs are best-characterized by a
single burst of age $t=8.25$ Gyr and metallicity $0.2$ solar.

Our stellar population synthesis analysis confirms
the results of visual inspections that the LRGs exhibit little recent
($t < 1$ Gyr) star formation activities. For both samples, the
best-fit models at $t \la 1.6$ Gyr are characterized by a single
burst of 2.5 solar metallicity.  Beyond 1.6 Gyr, the best-fit models have
$\tau$ ranging from $\tau=0.2$ to 0.5 Gyr.  The more extended $\tau$
models are compensated by the corresponding older age.  
At still older ages of $t>5$ Gyr, the best-fit models are either
single burst of 0.2 solar metallicity or a $\tau$ model of $\tau=0.5$
Gyr and solar metallicity.  The corresponding lower metallicity at
older ages is consistent with the well-known age-metallicity
degeneracy \citep{worthey1994a}, making it difficult to determine a
precise metallicity for the underlying stellar population.  However,
the results of our likelihood analysis shows that we can robustly
constrain the {\it minimum age} of the stellar population to be $> 1$
Gyr.

\section{Discussion} 

The results of the stellar population synthesis analysis can be
summarized as follows.  First, Mg\,II absorbing and non-absorbing LRGs
share comparable star formation histories, confirming that {\it the
  LRGs with associated Mg\,II do not constitute a biased population of
  star-forming galaxies}.  Second, these LRGs do not show recent star
formation activities.  Their mean spectra are best described by
evolved stellar populations of {\it at least} 1 Gyr old.

A caveat of our analysis is that if star formation occurs in the
outskirts of the LRGs excluded by the slit spectroscopy, we would not
have detected the blue light and emission lines associated with these
regions. In a recent analysis, \citet{tal2011a} have searched for
extended light around LRGs based on stacked SDSS images of more than
40000 objects. These authors find that while excess light can be
detected out to 100 kpc of the LRGs, the colors of the extended light
is consistent with the colors of the LRGs. The consistent colors
indicate that the stellar population does not vary significantly with
radius in the LRG halos.  Therefore we conclude that the observed
Mg\,II absorbers are unlikely to be associated with star-forming
regions in the outskirts of the LRGs.  Here we discuss the
implications of our analysis.

\begin{figure}
 \vspace{0.5pt}  \centerline{\hbox{ \hspace{0.0in}
	\includegraphics[angle=0,scale=0.30]{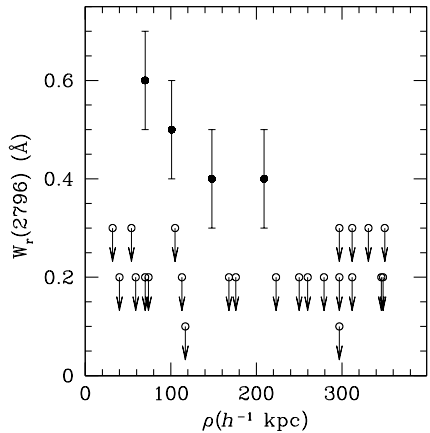}
      }
   }
\caption{ $W_r(2796)$ versus $\rho$ for the subsample of 28 LRGs. The absorbers strength 
of the four Mg\,II absorbers physically associated with LRGs are shown in solid black. The 2-$\sigma$ upper limits measured 
for the 24 non-absorbing LRGs are denoted by open circles and arrows. We obtain spectra of these 28 galaxies without prior 
knowledge of the presence/absence of Mg\,II absorbers in the QSO spectra (Sample B in G10). This allows us to measure a covering 
fraction of cool gas around LRGs. The covering fraction is $\kappa = 0.14 \pm 0.06 $ down to our 
detection limit of $W_r(2796)=0.3\,$\AA. 
}
\label{likelihood}
\end{figure}

\subsection{Transition in the cool gas content of dark matter halos}

In Figure 3, we present $W_r(2796)$ versus $\rho$ for a subsample of
LRGs. We obtained spectra of these LRGs without prior knowledge of the
presence/absence of Mg\,II absorbers in the QSO spectra (Sample B). 
Because these LRGs were selected with no
prior knowledge of Mg\,II absorbers, we can use this sample to measure
the covering fraction, $\kappa$, of cool gas in LRG halos. The
likelihood of detecting an ensemble of galaxies of which $n$ give rise
to Mg\,II absorbers and $m$ do not is given by \citep{chen2010a}
\begin{equation}
\mathcal{L}(\kappa) = \langle \kappa \rangle^n (1-\langle \kappa \rangle)^m  \; . 
\end{equation}
Within $\rho=350$ \hkpc four LRGs have associated Mg\,II absorbers
($n=4$) and 24 do not ($m=24$) show Mg\,II absorption features to
2-$\sigma$ upper limit better than $W_r(2796)=0.3$ \AA.  Using
equation (2), we found $\kappa = 0.14 \pm 0.06$, where the error bars
indicate the 68\% confidence interval based on the likelihood
function.  We also calculated $\kappa$ within the gaseous radius,
$R_{\rm gas}$, defined by \citet{chen2010a}. According to their model,
a $10^{13}$ \hmsol\ halo possess a gaseous envelope of size $R_{\rm
  gas} = 108$ \hkpc.  Within this gaseous radius, there are two
physical pairs and seven upper limits ($\kappa = 0.22 \pm 0.13$). 

Here we assume that a Mg\,II absorber found within $\rho=350$ \hkpc 
and $|\Delta v|=350$ \kms of an LRG traces gas that is physically associated 
with the LRG, i.e. the gas inhabits the dark matter halo of the LRG. The absorber 
could also trace gas in correlated large-scale galactic structures found 
in the volume probed by $\rho$ and $|\Delta v|$ and centered on the LRG. In this case, 
the gas would not be physically associated with the LRG. However, G10 
calculated the contribution of these correlated pairs for a sample 
of photometrically selected LRG--Mg\,II pairs and found the 
contamination level to be low ($\approx$ 3\% of the observed Mg\,II 
absorbers in the vicinity of LRGs are thought to arise from correlated structures). 
We thus considered the spectroscopic Mg\,II--LRG pairs to be physical pairs. 

We can compare the covering fraction measurements with the upper limits on $\kappa$ 
derived in G10. While these authors do not provide an upper limit on the covering 
fraction for absorbers with $W_r(2796)>0.3$\AA, they found that for the limiting 
strength of $W_r(2796)>0.5$\AA, $\kappa<0.18$. The upper limit derived for 
absorbers of $W_r(2796)>0.3$\AA\ is expected to be higher than the one 
derived for absorbers of $W_r(2796)>0.5$\AA. A value of $\kappa=0.14$ for 
absorbers with $W_r(2796)>0.3$\AA\ is thus consistent with the upper limits 
derived in G10.

The value of $\kappa$ in LRG halos is significantly smaller than
$\kappa \approx 0.7$ measured for $L*$ and sub-$L_*$ galaxies in
\citet{chen2010a} indicating a steep decline in the cool gas content
of dark matter halos.  In addition, we found that the cool gas
cross-section extends well beyond $R_{\rm gas}$. Such extended gaseous
envelope is not detected around low-mass halos \citep{chen2010a}.  The
sharp decline in the gas covering fraction is consistent with the
expectation of halo occupation distribution models of Tinker \& Chen
(2008, 2010) to explain the clustering amplitude of Mg\,II absorbers.
Such declining trend of cool gas fraction with halo mass is also
expected in theoretical models addressing the cool gas content of dark
matter halos (e.g.,\ \citealt{maller2004a,keres2009a}).

\subsection{Physical properties of cool clouds probed by Mg\,II}

The lack of recent star formation activity in the LRGs indicates
that the Mg\,II absorbing gas is likely due to infalling clouds
instead of starburst driven outflows.  Under the infall scenario,
different theoretical models have been studied to explain the presence
of cool halo gas.  These include cold flows (e.g.,\
\citealt{faucher-giguere2011a}), cool clouds condensed out of the hot
halos due to thermal instabilities (Mo \& Miralda-Escud\'e 1996;
Maller \& Bullock 2004, hereafter MB04) , and stripped material from
satellite galaxies (e.g.,\ \citealt{wang1993a,agertz2009a}).
Irrespective of the physical mechanism that produces the observed
Mg\,II absorbers near the LRGs, the cool clouds are expected to fall
and fuel star formation at the center of the halo.  At the same time,
the galaxies are found to be quiescent for at least the past 1 Gyr,
implying that these Mg\,II clouds do not survive the infall through
the hot halo.

Whether or not cool clouds can reach the center of the halo depends on
the infall time scale relative to the disruption time scale.  Under
hydrostatic equilibrium, the infall time scale is determined primarily
by the ram pressure drag, while the discruption time scale is driven
predominantly by heat conduction.
If the evaporation time scale, $\tau_{\rm evap}$, is shorter than the
time scale needed for ram pressure drag to sap the energy, the clouds
will likely evaporate before reaching the center of the halo. This
allows us to constrain the physical properties of the cool clouds
(e.g.\ Maller \& Bullock 2004).


For this particular model, $\tau_{\rm evap} \simeq
16\,m_6^{2/3}T_6^{-3/2}(\Lambda_z t_8)^{-1/3}~\rm{Gyr}$ where $m_6$ is
the cloud mass in units of $10^6$ $M_{\odot}$, $T_6 = T/10^6$\,K is
the halo gas temperature, $\Lambda_z$ and $t_8=t/(8~\rm{Gyr})$ are the
cooling parameter and halo formation time scale (see equation 35 and
Appendix B in MB04).  Assuming that the hot gas is isothermal with $T
\sim 6 \times 10^6$\,K, $t_8$ is the age of the universe at $z \sim
0.5$, and the metallicity of the gas is 0.1 solar, we found that
$\tau_{\rm evap} \simeq 1.1\,m_6^{2/3}~\rm{Gyr}$. The parameter
$\tau_{\rm evap}$ constitutes the typical time scale over which clouds
will be disrupted due to heat conduction.  
To calculate the infall time scale, we first adopted the maximum
rotation velocity of the halo to estimate the ram pressure drag force.
The time scale over which ram pressure drag force is then given by
$\tau_{\rm{rp}} \simeq 2.6\,m_6^{1/3} T_6^{-1/2} (\Lambda_z
t_8)^{1/3}~\rm{Gyr}$ (see equation 43 in MB04).
We found
$\tau_{\rm{rp}} \simeq 1.1\,m_6^{1/3}~\rm{Gyr}$.
Because the cool clouds are expected to be disrrupted before reaching
the center of the halo, $\tau_{\rm evap} \la \tau_{\rm rp}$. This
led to a mass limit of $m \la 10^6$\,M$_{\odot}$.


We note a main caveat in this calculation.  If the hot halo is out of
equilibrium either due to supernova driven wind or AGN outflows, then
the structure of halo gas distribution is expected to be different
before hydrostatic equilibrim can be restored (e.g.\ Brighenti \&
Mathews 2003).  The estimated mass limit is therefore very uncertain,
depending on the thermal state of the halo.  For example,
outflows/winds are expected to increase the ram pressure, prolonging
the infall time, and thereby increasing the mass limit.



It is also possible that the observed Mg\,II absorbers originate in
the ISM of a satellite galaxy, directly intercepting the QSO
sightline.  However, we find this an unlikely scenario.
\citet{gauthier2010a} have shown based on an analytic calculation that
if the gas content of satellite galaxies remains intact as they orbit
around the primary galaxy, then satellite galaxies could be a dominant
contributor to the gas cross-section only at small distances ($\rho <
100$ \hkpc).  Given that ram pressure and tidal stripping is effective
in removing gas in satellite galaxies (e.g.\
\citealt{chynoweth2008a}), we consider this a conservative upper limit
to possible satellite contributions to the observed Mg\,II absorption
features.  We expect that empirical knowledge of the satellite
environment of the LRGs and gas kinematics of the absorbers will
provide further insights into the physical origin of the observed
extended gas in LRG halos.


\section*{Acknowledgments} 

We thank Andrey Kravtsov, Mariska
Kriek, Michael Rauch, and Vivien Wild for illuminating discussions.
We also thank Don York for helpful comments on an earlier draft of the
paper. We would like to thank the anonymous referee for their insightful 
comments that improved the draft significantly. JRG acknowledges 
support from the Brinson Predoctoral Fellowship and by a 
Grant-In-Aid of Research from the National Academy
of Sciences, administered by Sigma Xi, The Scientific Research
Society.


\begin{thebibliography}{24}
\expandafter\ifx\csname natexlab\endcsname\relax\def\natexlab#1{#1}\fi

\bibitem[{{Agertz} {et~al}(2009)}]{agertz2009a}
{Agertz}, O., {Teyssier}, R., \& {Moore}, B. 2009, MNRAS, 397, L64


\bibitem[{{Birnboim} \& {Dekel}(2003)}]{birnboim2003a}
{Birnboim}, Y., \& {Dekel}, A. 2003, MNRAS, 345, 349

\bibitem[{{Blake} {et~al.}(2007){Blake}, {Collister}, {Bridle}, \&
  {Lahav}}]{blake2007a}
{Blake}, C., {Collister}, A., {Bridle}, S., \& {Lahav}, O. 2007, MNRAS, 374,
  1527

\bibitem[{{Blake} {et~al.}(2008){Blake}, {Collister}, \& {Lahav}}]{blake2008a}
{Blake}, C., {Collister}, A., \& {Lahav}, O. 2008, MNRAS, 385, 1257

\bibitem[{{Bowen} \& {Chelouche}(2011)}]{bowen2011a}
{Bowen}, D.~V., \& {Chelouche}, D. 2011, ApJ, 727, 47

\bibitem[{{Brighenti} \& {Mathews}(2003)}]{brighenti2003a}
{Brighenti}, F., \& {Mathews}, W.G. 2003, ApJ, 587, 580

\bibitem[{{Bruzual} \& {Charlot}(2003)}]{bruzual2003a}
{Bruzual}, G., \& {Charlot}, S. 2003, MNRAS, 344, 1000

\bibitem[{{Chabrier}(2003)}]{chabrier2003a}
{Chabrier}, G. 2003, PASP, 115, 763

\bibitem[{{Charlot} \& {Fall}(2000)}]{charlot2000a}
{Charlot}, S., \& {Fall}, M.S. 2000, ApJ, 539, 718

\bibitem[{{Chen} {et~al.}(2010){Chen}, {Helsby}, {Gauthier}, {Shectman},
  {Thompson}, \& {Tinker}}]{chen2010a}
{Chen}, H., {Helsby}, J.~E., {Gauthier}, J., {Shectman}, S.~A., {Thompson},
  I.~B., \& {Tinker}, J.~L. 2010, ApJ, 714, 1521

\bibitem[{{Chynoweth} {et~al.}(2008){Chynoweth}, {Langston}, {Yun}, {Lockman},
  {Rubin}, \& {Scoles}}]{chynoweth2008a}
{Chynoweth}, K.~M., {Langston}, G.~I., {Yun}, M.~S., {Lockman}, F.~J., {Rubin},
  K.~H.~R., \& {Scoles}, S.~A. 2008, AJ, 135, 1983
  
\bibitem[{{Ciotti} \& {Ostriker}(2007)}]{ciotti2007a}
{Ciotti}, L., \& {Ostriker}, J.P. 2007, ApJ, 665, 1038

\bibitem[{{Collister} {et~al.}(2007){Collister}, {Lahav}, {Blake}, {Cannon},
  {Croom}, {Drinkwater}, {Edge}, {Eisenstein}, {Loveday}, {Nichol}, {Pimbblet},
  {de Propris}, {Roseboom}, {Ross}, {Schneider}, {Shanks}, \&
  {Wake}}]{collister2007a}
{Collister}, A., {et~al.} 2007, MNRAS, 375, 68

\bibitem[{{Faucher-Giguere} {et~al.}(2011){Faucher-Giguere}, {Keres}, \&
  {Ma}}]{faucher-giguere2011a}
{Faucher-Giguere}, C., {Keres}, D., \& {Ma}, C. 2011, ArXiv e-prints

\bibitem[{{Gauthier} {et~al.}(2009){Gauthier}, {Chen}, \&
  {Tinker}}]{gauthier2009a}
{Gauthier}, J.-R., {Chen}, H.-W., \& {Tinker}, J.~L. 2009, ApJ, 702, 50

\bibitem[{{Gauthier} {et~al.}(2010){Gauthier}, {Chen}, \&
  {Tinker}}]{gauthier2010a}
---. 2010, ApJ, 716, 1263

\bibitem[{{Kacprzak} {et~al.}(2010){Kacprzak},{Churchill},{Ceverino},
{Steidel},{Klypin}, \& {Murphy}}]{kacprzak2010a} {Kacprzak}, G.G., {et~al}, 2010, ApJ, 711, 533 

\bibitem[{{Kere{\v s}} {et~al.}(2009){Kere{\v s}}, {Katz}, {Fardal},
  {Dav{\'e}}, \& {Weinberg}}]{keres2009a}
{Kere{\v s}}, D., {Katz}, N., {Fardal}, M., {Dav{\'e}}, R., \& {Weinberg},
  D.~H. 2009, MNRAS, 395, 160

\bibitem[{{Magorrian} {et~al.}(1998){Magorrian},{Tremaine},{Richstone},
{Bender},{Bower},{Dressler},{Faber},{Gebhardt},{Green},{Grillmair},{Kormendy}, 
\& {Lauer}}]{magorrian1998a} {Magorrian}, J., {et~al}, 1998, ApJ, 115, 2285 

\bibitem[{{Maller} \& {Bullock}(2004)}]{maller2004a}
{Maller}, A.~H., \& {Bullock}, J.~S. 2004, MNRAS, 471

\bibitem[{{Marigo} \& {Girardi}(2007)}]{marigo2007a}
{Marigo}, P., \& {Girardi}, L. 2007, AAP, 469, 239

\bibitem[{{Mo} \& {Miralda-Escude}(1996)}]{mo1996a}
{Mo}, H.~J., \& {Miralda-Escude}, J. 1996, ApJ, 469, 589

\bibitem[{{Roseboom} {et~al.}(2006){Roseboom}, {Pimbblet}, {Drinkwater},
  {Cannon}, {de Propris}, {Edge}, {Eisenstein}, {Nichol}, {Smail}, {Wake},
  {Bland-Hawthorn}, {Bridges}, {Carson}, {Colless}, {Couch}, {Croom}, {Driver},
  {Hewett}, {Loveday}, {Ross}, {Schneider}, {Shanks}, {Sharp}, \&
  {Weilbacher}}]{roseboom2006a}
{Roseboom}, I.~G., {et~al.} 2006, MNRAS, 373, 349

\bibitem[{{Schneider} {et~al.}(2007){Schneider}, {Hall}, {Richards}, {Strauss},
  {Vanden Berk}, {Anderson}, {Brandt}, {Fan}, {Jester}, {Gray}, {Gunn},
  {SubbaRao}, {Thakar}, {Stoughton}, {Szalay}, {Yanny}, {York}, {Bahcall},
  {Barentine}, {Blanton}, {Brewington}, {Brinkmann}, {Brunner}, {Castander},
  {Csabai}, {Frieman}, {Fukugita}, {Harvanek}, {Hogg}, {Ivezi{\'c}}, {Kent},
  {Kleinman}, {Knapp}, {Kron}, {Krzesi{\'n}ski}, {Long}, {Lupton}, {Nitta},
  {Pier}, {Saxe}, {Shen}, {Snedden}, {Weinberg}, \& {Wu}}]{schneider2007a}
{Schneider}, D.~P., {et~al.} 2007, AJ, 134, 102

\bibitem[{{Stewart} {et~al.}(2010){Stewart}, {Kaufmann}, {Bullock}, {Barton},
  {Maller}, {Diemand}, \& {Wadsley}}]{stewart2010a}
{Stewart}, K.~R., {Kaufmann}, T., {Bullock}, J.~S., {Barton}, E.~J., {Maller},
  A.~H., {Diemand}, J., \& {Wadsley}, J. 2010, ArXiv e-prints

\bibitem[{{Tal} \& {van Dokkum}(2011)}]{tal2011a}
{Tal}, T., \& {van Dokkum}, P. 2011, ArXiv e-prints

\bibitem[{{Tojeiro} {et~al.}(2011){Tojeiro},{Percival},{Heavens}, \&{Jimenez}}]{tojeiro2011a}
{Tojeiro}, R., {Percival}, W.J., {Heavens}, A.F., \& {Jimenez}, R. 2011, MNRAS, 413, 434

\bibitem[{{Thomas} {et~al.}(2005){Thomas}, {Maraston}, {Bender}, \& {Mendes de
  Oliveira}}]{thomas2005a}
{Thomas}, D., {Maraston}, C., {Bender}, R., \& {Mendes de Oliveira}, C. 2005,
  ApJ, 621, 673

\bibitem[{{Tinker} \& {Chen}(2008)}]{tinker2008a}
{Tinker}, J.~L., \& {Chen}, H.-W. 2008, ApJ, 679, 1218

\bibitem[{{Tinker} \& {Chen}(2010)}]{tinker2009b}
---. 2010, ApJ, 709, 1

\bibitem[{{Wang}(1993)}]{wang1993a}
{Wang}, B. 1993, ApJ, 415, 174

\bibitem[{{Westmeier} {et~al.}(2007){Westmeier}, {Braun}, {Br{\"u}ns}, {Kerp},
  \& {Thilker}}]{westmeier2007a}
{Westmeier}, T., {Braun}, R., {Br{\"u}ns}, C., {Kerp}, J., \& {Thilker}, D.~A.
  2007, 51, 108

\bibitem[{{Wild} {et~al.}(2010)}]{wild2010a}
{Wild}, V., {Heckman}, T., \& {Charlot}, S. 2010, MNRAS, 405, 933

\bibitem[{{Worthey}(1994)}]{worthey1994a}
{Worthey}, G. 1994, ApJS, 95, 107

\bibitem[{{Worthey} {et~al.}(2011)}]{worthey2011a}
{Worthey}, G., {Ingermann}, B.A., \& {Serven}, J. 2011, ApJ, 729, 148
\end{thebibliography}

\begin{table*}
 \centering
 \begin{minipage}{180mm}
 \caption{Summary of the Absorbing and Non-Absorbing Samples of LRGs}
 \begin{tabular}{@{}cccccccccc@{}}
 \hline
ID  & $z_{\rm spec}$ & $i'$ & Instrument & Exptime (sec) & UT date & $W_r(2796)$ & 
$\rho (\hkpc)$ & $|\Delta v|$ (\kms) & Sample \\
(1) & (2) & (3) & (4) & (5) & (6) & (7) & (8) & (9) & (10)  \\
\hline
 \multicolumn{9}{c}{\textsc{Absorbing Galaxies}} \\
\hline 
SDSSJ114445.36$+$071456.4 & 0.4906 & 19.26 & B\&C & 2700 & 2010-04-15 & $0.6\pm0.1$ & { }70 & 380 & B \\
SDSSJ114658.60$+$020716.8 & 0.5437 & 19.50 & DIS & 1800$+$1500& 2010-03-12 & $1.6\pm0.2$ & { }54 & 198 & A  \\
SDSSJ142242.72$+$041512.0 & 0.5512 & 19.85 & DIS & 2$\times$1800 & 2011-03-15 & $0.4\pm0.1$ & 148 & { }30 & B\\
SDSSJ150638.16$+$041906.9 & 0.6155 & 19.69 & B\&C & 2400$+$2400 & 2010-04-12 & $0.4\pm0.1$ & 209  & { }20 & B  \\
SDSSJ160725.87$+$471221.7  & 0.4980 & 19.70 & DIS & 2$\times$1800 & 2009-05-26 & $1.2\pm0.2$  & 134 & {   }0  & A \\
SDSSJ161713.68$+$243254.0 & 0.5703 & 19.06 & DIS & 2$\times$1800 & 2009-05-26 & $1.5\pm0.3$ & { }33 &  126 & A \\
SDSSJ211625.92$-$062415.4  & 0.5237 & 19.31 & B\&C & 2$\times$2400$+$2000 & 2009-09-21 & $0.5\pm0.1$  & 101 & { }79 & B   \\
SDSSJ220703.36$-$090051.6 & 0.5604 & 19.60 & DIS & 2$\times$1800 & 2010-07-20 & $4.0\pm0.1$ & 171 & 360 & A \\[2mm]
\hline \\[-4mm]
\multicolumn{9}{c}{\textsc{Non-absorbing Galaxies}} \\[1mm]
\hline \\
SDSSJ001150.19$+$160434.7 & 0.5321 & 19.09 & B\&C & 2$\times$2400 & 2010-09-04 & $<$0.2 & 250  & - & B \\ 
SDSSJ003344.50$-$005459.8 & 0.5019 & 19.644 & DIS &2$\times$1800 &  2010-09-11 & $<$0.3 & 297 & - & B\\
SDSSJ005759.11$+$152013.6 & 0.5347 & 18.95 & DIS & 2$\times$1800  & 2010-12-23 & $<$0.2 &  279 & - & B \\
SDSSJ010543.69$+$004121.9 & 0.5356 & 19.13 & B\&C & 2$\times$2400 & 2010-09-04 & $<$0.2 & 346 & - & B  \\
SDSSJ012414.83$+$145510.6 & 0.5742 & 19.74 & DIS & 2$\times$1800  & 2010-09-16 & $<$0.3 & { }54  & - & B \\
SDSSJ014728.03$+$142231.8 &  0.6500 & 19.70 & DIS & 2$\times$2400 & 2010-12-23 & $<$0.2 & 223 & - & B\\
SDSSJ015452.46$-$095533.6  & 0.5663 & 19.98 & B\&C & 2$\times$2400 & 2009-09-21 & $<$0.2 & { }40  & - & B\\
SDSSJ020107.70$+$125812.4 & 0.6259 & 19.52 & DIS & 2$\times$1800 & 2010-09-11 & $<$0.3 & 310 & - & A \\
SDSSJ034802.50$-$070339.3  & 0.6105 & 19.08 & B\&C & 2$\times$2400 & 2009-09-18 & $<$0.3 & 199 & - & A \\
SDSSJ082336.96$+$064436.2 & 0.5347 & 19.40 & DIS & 2$\times$1800 & 2011-01-03 & $<$0.3 &  { }32 & - & B \\
SDSSJ100709.60$+$043133.7 & 0.5783 & 19.29 & B\&C & 2$\times$2700$+$1800 & 2010-04-19 & $<$0.1 & 297  & - &  B\\
SDSSJ104934.80$+$075750.1 & 0.4793 & 19.56 & DIS & 2$\times$1800 & 2010-03-13 & $<$0.2 & 123 & - & A \\
SDSSJ121357.60$+$022718.2 & 0.5263 & 18.24 & B\&C & 2400 & 2010-04-14 & $<$0.3 & 312 & - & B \\
SDSSJ125300.00$+$005429.6 & 0.5402 & 19.07 & SDSS & -  & - & $<$0.3 & 350 & - & B\\
SDSSJ131200.00$+$013413.6 & 0.5425 & 19.10 & B\&C & 2$\times$2400 & 2010-04-18 & $<$0.3 & 312 & - & B\\
SDSSJ131815.84$+$012437.9 & 0.5405 & 20.08 & DIS & 2$\times$1800 & 2010-06-14 & $<$0.2 & { }74  & - & B \\
SDSSJ133637.20$+$024130.4 & 0.5936 & 19.12 & B\&C & 2$\times$2700 & 2010-04-15 & $<$0.2 & 176 & - & B \\
SDSSJ140215.12$+$070946.4 & 0.6682 & 20.00 & DIS & 2$\times$1800 & 2010-06-05 & $<$0.3 & 105  & - & B \\
SDSSJ141004.08$+$064352.5 & 0.4830 & 19.74 & DIS & 2$\times$1800 & 2010-03-18 & $<$0.2 & 246 & - & A \\
SDSSJ141525.92$+$132929.8 & 0.5375 & 19.48 & DIS & 2$\times$1800 & 2010-06-05 & $<$0.2 & 168  & - & B \\
SDSSJ141654.48$-$000534.0 & 0.4746 & 19.39 & B\&C &2700$+$300 & 2010-04-15 & $<$0.2 & { }59 & - & B \\
SDSSJ142312.00$+$093409.5 & 0.6139  & 19.06 & B\&C & 2400$+$2100 & 2010-04-12 & $<$0.1  & 117 & - & B \\
SDSSJ153001.68$-$012535.9 & 0.5830 & 19.94 & B\&C & 2700$+$1000 & 2010-04-14 & $<$0.2 & 297& - & B \\
SDSSJ155445.60$+$084102.8 & 0.4951 & 19.75 & B\&C &2$\times$2700 & 2010-04-15 & $<$0.2  & 348 & - & B  \\
SDSSJ160954.48$+$065513.6 & 0.5308  & 19.48 & B\&C & 2$\times$2700 & 2010-04-12 & $<$0.3 & 322 & - & A \\
SDSSJ214806.72$-$004436.1 & 0.5443 & 19.57 & B\&C & 2$\times$2400 & 2010-09-04 & $<$0.3 & 331& - & B \\
SDSSJ232737.57$+$153324.1 & 0.4756 & 20.02 & DIS & 2400$+$1200 & 2010-10-07 & $<$0.2 & 113& - & B \\
SDSSJ232924.13$-$100728.9 & 0.4606 & 19.35 & B\&C & 2$\times$2400 & 2009-09-05 & $<$0.2 &  { }70 & - &  B \\
\hline
\end{tabular}
\medskip
For non-absorbing galaxies, $W_r(2796)$ is a 2-$\sigma$ upper-limit on Mg\,II $\lambda$2796 transition strength.
$|\Delta v|$ is the absolute velocity separation between the Mg\,II absorber and the LRG redshift.
\end{minipage}
\end{table*}

\end{document}